\documentstyle[12pt]{article}
\newcommand{\bb}{\begin{eqnarray}}
\newcommand{\ee}{\end{eqnarray}}
\begin{document}
\title{{A note on the entropy of charged \\ multi - black - holes
}}
\author{P. Mitra\thanks{e-mail mitra@tnp.saha.ernet.in}\\
Saha Institute of Nuclear Physics\\
Block AF, Bidhannagar\\
Calcutta 700 064, INDIA}
\date{gr-qc/9908014}
\maketitle
\begin{abstract}
Majumdar - Papapetrou multi - black - hole solutions of the Einstein - Maxwell equations
are considered in four and higher dimensions. The Euclidean action with 
boundary conditions appropriate to the canonical ensemble is shown to lead to 
zero entropy. 
\end{abstract}

\bigskip
\section{Introduction}
A field configuration given by
\bb
ds^2=-H^{-2}dt^2+H^2 (dr^2+r^2d\Omega^2),~~
A_t=H^{-1},
\ee
with $H(\vec r)$ satisfying 
\bb
\nabla^2 H=0,
\ee 
represents a (static) solution of the Einstein - Maxwell equations.
A general form for $H$ with only point singularities is
\bb
H= {\rm const}+\sum_a {c_a\over |\vec r-\vec r_a|},
\ee
and it can be
interpreted as being composed of a set of extremal charged black holes. In fact,
the special case in which the sum has only one term can be converted to the
standard extremal Reissner - Nordstr\"{o}m metric by a translation of $\vec r$. 
A collection of black holes does not usually yield a static configuration,
but in this case of extremal charged black holes, the
gravitational and electrostatic forces cancel pairwise, as one can expect from
a Newtonian system of appropriately charged particles. Spacetimes of this
kind were discovered over fifty years back \cite{majum} and have been analysed
in \cite{HP} and other papers.

Now black holes are assigned temperatures and entropies in quantum theory, at least
in a semiclassical approach. It is natural to wonder if the thermodynamical
properties of these composite systems are similar to those of an extremal
charged black hole.   Unlike an ordinary black hole, charged or uncharged, an
extremally charged black hole exhibits special properties. Whereas the surface
gravity gives a measure of the temperature of a black hole in ordinary
circumstances, this quantity vanishes for an extremally charged black hole.
Thus such black holes have sometimes been assigned a zero temperature.
Euclidean semiclassical gravity however suggests that the temperatures of such
black holes should be regarded as arbitrary \cite{HHR}. This is related to the
fact that the Euclidean versions of these metrics do not exhibit the conical
singularity that Euclidean versions of metrics of ordinary black hole
spacetimes show. The behaviour of the entropy too can be different.  Ordinarily
a black hole is assigned an entropy equal to a quarter of its horizon area in
natural units. The approach in which an extremal black hole is considered to
have an arbitrary temperature suggests on the other hand a zero entropy for
such black holes \cite{HHR}. This would indicate that extremal black holes are
special objects not continuously related to non-extremal black holes.
However, this is not the complete picture. It is best to say that there
are two inequivalent approaches to the study of extremal charged black holes
\cite{GM}. There is the approach of starting with a classical extremal
black hole and proceeding to quantize it with fixed topology. This leads to an
arbitrary temperature and a zero entropy. The alternative is to start with a
quantum black hole and fix its properties by demanding that it be extremal.
This leads to continuity: a zero temperature and an entropy given by a quarter of
the horizon area.  These alternatives have been shown to occur for
asymptotically flat Reissner - Nordstr\"{o}m black holes \cite{GM} as well as
asymptotically anti-de Sitter ones \cite{ads, ads-c}.

In view of this situation, it is natural to ask
what happens in the case of a multi - black hole spacetime, which is, 
as we have seen, a generalization of an extremal charged black hole.  
Can both of these two alternatives be implemented here,
so that the temperature can be zero or arbitrary and the
entropy can equal a quarter of the horizon area or vanish?  
This is the question to be studied here.
But before going into this question, a brief discussion of multi - black
- hole solutions in higher dimensions will be presented.

\section{Multi - black - holes in higher dimensions}
A cancellation of gravitational and electrostatic forces can be expected to
occur in higher dimensions too, but if the above metric is directly transcribed
in higher dimensions, it fails to satisfy the field equations even if $H$ is
a harmonic function. A minor adjustment is needed,
and the proper generalization of the above configuration is 
\bb
ds^2=-H^{-2}dt^2+H^{2\over d-3} (dr^2+r^2d\Omega_{d-2}^2),~~
A_t=\sqrt{d-2\over 2(d-3)}H^{-1}.
\ee
This is a solution of the $d$-dimensional Einstein - Maxwell equations if 
$H$ is again a harmonic function, for example of the form
\bb
H={\rm const}+\sum_a {c_a\over |\vec r-\vec r_a|^{d-3}}.\label{H}
\ee
This represents again a collection of several extremal charged black holes.

The Euclidean version of this $d$-dimensional configuration  is
\bb
ds^2=H^{-2}d(x^0)^2+H^{2\over d-3} (dx^idx^i),~~
A_0=i\sqrt{d-2\over 2(d-3)}H^{-1},~~A_i=0.\label{g}
\ee
The metric (\ref{g}) gives rise to an Einstein tensor
\bb
E_{00}&=& {d-2\over d-3} H_{,kk}H^{-{3d-7\over d-3}}-{1\over 2} {d-2\over d-3}
H_{,k}H_{,k}H^{-{4d-10\over d-3}},
\nonumber\\
E_{0i}&=&0,\nonumber\\
E_{ij}&=&-{d-2\over d-3} H_{,i}H_{,j}H^{-2}+{1\over 2} {d-2\over d-3}
H_{,k}H_{,k}H^{-2}\delta_{ij}.
\ee
The interesting fact here is that the second derivatives of $H$ enter this tensor only
through $E_{00}$ and in the combination $H_{kk}$. The same is true of 
$F^{\mu\nu}_{;\nu}$ if one chooses $A_i$ to be zero and $A_0$ to be a function of $H$.
In fact, if one takes the normalization
\bb
A_0&=&\pm i\sqrt{d-2\over 2(d-3)}H^{-1},
\ee
one has
\bb
E_{\mu\nu}=2(F_{\mu\tau}{F_\nu}^\tau -{1\over 4}g_{\mu\nu}F_{\rho\sigma}
F^{\rho\sigma}),
\ee
and also
\bb
F^{\mu\nu}_{;\nu}=0,
\ee
provided that
\bb
H_{,kk}=0,
\ee
so that the Einstein - Maxwell equations are satisfied if $H$ is a harmonic function.
The general solution for $H$ with only point singularities 
was given in (\ref{H}) and 
can be interpreted as a set of extremal black holes.

This solution has a multi-component horizon given by $\vec r=\vec r_a$, corresponding to
the singularities of $H$. Although this seems to be a set of points, the $(d-2)$
dimensional area indicated by the metric (\ref{g}) is finite if calculated by a
limiting procedure  $\vec r\to\vec r_a$ for each $a$. 
The total area is given by
\bb
A_{d-2}\sum_a c_a^{d-2\over d-3},
\ee 
where $A_{d-2}$ is the area of a unit sphere
in $(d-2)$ dimensions with the {\it usual} Euclidean metric.

One can check that there is no conical singularity about $\vec r=\vec r_a$ - 
this means that the Euclidean time can be compactified with an arbitrary
period and the temperature may therefore be regarded as arbitrary.

\section{Action and entropy}

The usual (grand canonical) action for this Euclidean system is
\bb
-{1\over 16\pi}\int_{\cal M} d^dx\sqrt gR -{1\over 8\pi}\int_{\partial{\cal M}}  
d^{d-1}x\sqrt h(K-K_0) +{1\over 16\pi}\int_{\cal M}d^dx\sqrt g F_{\mu\nu}F^{\mu\nu},
\ee
where a manifold ${\cal M}$ with a boundary is envisaged. The boundary will ultimately
be taken to infinity, but as long as it is there, the boundary term ensures that
the variation of this action with fixed metric on the boundary reproduces the Einstein
equations. $K$ is the extrinsic curvature of the boundary and $K_0$ the same object
calculated in the usual flat metric. 
The Maxwell equations are also reproduced with fixed potential $A_0$ on the
boundary, which is why this is to be thought of as the grand canonical action. 
Semiclassical studies on black hole entropy are usually carried out
in this grand canonical formulation. However, the extremal Reissner - Nordstr\"{o}m
black hole is not a minimum of the grand canonical action \cite{GMcom}. It is
necessary to use the canonical formulation. There is a problem in defining
the canonical partition function for uncharged or weakly charged black holes,
as the integral over energies appears to diverge, but this problem does not occur
for sufficiently charged black holes, in particular for near - extremal ones.
The extremal black hole is indeed a minimum of the action corresponding to
the canonical ensemble, as was first seen in the context of an asymptotically
anti-de Sitter charged black hole, and the result continues to hold in the limit
of zero cosmological constant 
\cite{ads-c}. Strictly speaking, this is known to be the case for a single
extremal black hole, but that is motivation enough to use the canonical ensemble for
multi - black holes.
 
To change to the canonical ensemble, that is, to alter
the boundary condition for the Maxwell equations to be of the form of fixed $F^{i0}$,
{\it i.e.,} of fixed charge, a boundary piece has to be added to the action:
\bb
-{1\over 4\pi}\int_{\partial{\cal M}} d^{d-1}x\sqrt h n_\mu F^{\mu\nu}A_\nu.
\ee

To calculate the on-shell action, one may simplify the combination of the Einstein
and the Maxwell terms by noting the relation
\bb
R={4-d\over 2-d}F_{\mu\nu}F^{\mu\nu}
\ee
which follows from the Einstein equation. Hence the combination becomes
\bb
-{2\over 4-d}{1\over 16\pi}\int d^dx \sqrt {g}R&=&
{1\over 8\pi(4-d)}\int d^dx H^{2\over d-3} H^{-{2d-4\over d-3}} H_{,i}H_{,i}{d-4
\over d-3}\nonumber\\ &=&
-{\beta\over 8\pi(d-3)}\int d^{d-1}xH_{,i}H_{,i}H^{-2}\nonumber\\ &=&
{\beta\over 8\pi(d-3)}\int d^{d-1}x\partial_i(H_{,i}H^{-1})\nonumber\\ &=&
{\beta\over 8\pi(d-3)}\int d^{d-2}xn_iH_{,i}H^{-1}\nonumber\\ &=&
-{\beta\over 8\pi(d-3)}A_{d-2}\sum_a c_a(d-3)A_{d-2}\nonumber\\ &=&
-{\beta\over 8\pi}A_{d-2}\sum_a c_a.
\ee
Here the $x_0$ integral has been replaced by $\beta$, and the space integral has been
converted into a surface integral by an integration by parts. The surface has been taken
to infinity, as indicated earlier, through the form $r=$constant. 
$n_i$ is normal to the surface and a unit vector in the {\it flat metric}. 
 
To calculate the $K$ terms, one first notes that the normal to the surface has the
radial component
\bb
n^r={1\over\sqrt{g_{rr}}}=H^{-{1\over d-3}}.
\ee
Hence,
\bb
K=(\Gamma^\mu_{r\mu}-\Gamma^r_{rr})n^r=\bigg({2\over d-3}H_{,r}H^{-1}+{d-2\over r}-
{1\over d-3}H_{,r}H^{-1}\bigg)H^{-{1\over d-3}}.
\ee
The integral of this quantity diverges because of the $1/r$ piece, but it gets removed
when $K_0$ is subtracted: $H$ can be replaced by unity for large $r$. Hence,
\bb
-{1\over 8\pi}\int d^{d-1}x\sqrt{h}(K-K_0)&=&
-{\beta\over 8\pi}\int d^{d-2}x H^{1\over d-3}
{1\over d-3} H_{,r}H^{-{d-2\over d-3}}\nonumber\\ &=&
{\beta\over 8\pi}{1\over d-3} (d-3) A_{d-2}\sum_a c_a\nonumber\\ &=&
{\beta\over 8\pi}A_{d-2}\sum_a c_a.
\ee
Lastly, the piece added to convert the grand canonical action to a
canonical ensemble action can be written as
\bb
-{\beta\over 4\pi}\int d^{d-2}x H^{1\over d-3} A_0\partial_i A_0 H^{2d-8\over d-3}
H^{-{1\over d-3}}n_i\nonumber\\ &=&
{\beta\over 4\pi}{1\over 2}{(d-2)\over(d-3)}\sum_a c_a(d-3)A_{d-2}\nonumber\\ &=&
{\beta\over 8\pi}(d-2)A_{d-2}\sum_a c_a.
\ee

Thus the on-shell action reduces to
\bb
{\beta\over 8\pi}(d-2)A_{d-2}\sum_a c_a,
\ee
which is proportional to $\beta$, indicating that the entropy must be
\bb
S=0.
\ee
Exactly as in the case of a single extremal charged black hole in four dimensions,
the temperature is arbitrary and the entropy zero. 

\section{Conclusion}

The approach of quantizing a classical black hole with definite topology has
been followed here. What is the alternative? One would have to start with a
quantum black hole where the topology has not been fixed at the classical level
and impose the extremal condition after identifying the minimum action
configuration as demanded by the lowest order semiclassical approximation. In
the investigations on single black holes \cite{GM, ads, ads-c}, one had to
consider the competition between solutions of the equations of motion with
extremal topology and solutions with non-extremal topology. In the present
situation however, solutions of the equations of motion with non-extremal
topology are not known, except in the special case of a single extremal black
hole. One can certainly consider {\it configurations} of non-extremal topology,
but they will not be at minima of the action because they are not solutions of the
equations of motion. Consequently it does not appear that the approach of
extremalization after quantization can lead to a different action or different
entropy. In other words, at least for the time being, one has only one
possibility: arbitrary temperature and zero entropy.

\end{document}